\documentclass[fleqn,12pt,twoside]{article}
\usepackage{espcrc1}
\usepackage{graphicx}
\def\no{\noindent}
\def\bc{\begin{center}}
\def\ec{\end{center}}

\def\beq{\begin{equation}}
\def\eeq{\end{equation}}

\def\Tr{\rm Tr}

\begin{document}

\title{Mixtures of fermionic atoms in an optical lattice}

\author{K. Ziegler\\
Institut f\"ur Physik, Universit\"at Augsburg\\
D-86135 Augsburg, Germany
}


\maketitle

A mixture of light and heavy spin-polarized fermionic atoms in an
optical lattice is considered.
Tunneling of the heavy atoms is neglected such that they are only
subject to thermal fluctuations. This results in a complex interplay 
between light and heavy atoms caused by quantum tunneling of the light
atoms. The distribution of the heavy atoms is studied. It can be 
described by an Ising-like distribution with a first-order
transition from homogeneous to staggered order. The latter is caused by
an effective nonlocal interaction due to quantum tunneling of the light atoms.
A second-order transition is also possible between an ordered
and a disordered phase of the heavy atoms. 

\section{Introduction: Phase separation and the physics of correlated disorder}

Mixtures of different species of cold atoms open an interesting field
of many-body physics, where the competition between the atoms 
may lead to new quantum states. This is the case for low-temperature
properties (i.e. for the ground state) as well as in the presence of
thermal fluctuations. For the latter case we expect phase separation, where
different phases can coexist. This phenomenon has been intensively discussed
in the solid-state community in the context of complex materials
\cite{littlewood}, and it is believed to be a result of the competition of different
(e.g. spin and orbital) degrees of freedom \cite{dagotto}. A similar situation may
appear in atomic mixtures due to correlations caused by the competition of the different
atomic species. Two or more phases can coexist if there is a first-order phase transition
between the individual phases. In other words, thermal fluctuations lead to metastable states
that decay only on long-time scales. Such a phenomenon is characterized by a broken
translational invariance. In terms of an atomic mixture in an optical lattice
the atoms are arranged in clusters, consisting of one of the phases, with a characteristic 
length scale relative to the optical lattice constant. This implies a specific type of 
correlated disorder that has interesting dynamical properties of the mixtures, 
including the localization of atoms. 

The physics of disorder in statistical or solid-state physics 
is associated with random scattering of light particles
by a {\it disordered} (i.e. a randomly produced)
array of slow particles. This disordered array is static and does not change
during the course of the considered scattering process. For instance, the scattering of
electrons in a crystal with randomly distributed impurity atoms is a 
typical realization of disordered physics or the scattering of photons in atmospheric dust. 

A comparable situation is found in an atomic system subject to an optical lattice. 
The periodic structure of the
optical lattice, given by counterpropagating lasers, is normally
quite robust such that disorder in the sense of impurities cannot
easily be achieved. However, in contrast to a crystal in solid-state
physics, two or more different types of atoms can be mixed and
brought in an optical lattice, where one type of the atoms is
relatively light (e.g. $^6$Li) and the other type(s) is (are)
heavy (e.g. $^{40}$K, $^{23}$Na \cite{ketterle04}, 
$^{87}$Rb). The different masses lead to different
dynamical properties. In particular, heavy atoms behave almost
like static degrees of freedom, in comparison to the light atoms.
Therefore, heavy atoms play the role of the impurity atoms in the
crystal and the light atoms the role of the electrons. Thus disorder
physics can be studied in mixtures of atoms with different masses which
are available in experiments \cite{demarco99}. The main difference is
here that we shall study the interaction of two species of atoms
within the same model. In other words, the distribution of disorder
(i.e. the distribution of the heavy atoms) is a direct consequence of
the interaction between light and heavy atoms. 

Disordered systems have attracted a lot of attention because they
provide a new class of physics, including new phases and new types of 
phase transitions. The main reason is that scattering of quantum
particles by a periodic structure is qualitatively different from
scattering by a disordered structure: The former 
is governed by Bloch's Theorem \cite{bloch28}
which states that the scattered quantum states are not strongly
affected by the scattering process but keep their wave-like
properties of the unscattered states. This changes dramatically
as soon as the periodicity of the scatterers is disturbed.
At first the quantum states loose their wave-like character because
the physics is controlled by diffusion. And for stronger disorder
the quantum states can even be localized \cite{anderson58}. There
is a phase transition between the diffusion-controlled regime
and the localized regime \cite{abraham79}.

A fundamental concept for a theoretical description of
disordered system is that physical quantities are averaged
with respect to a statistical distribution of the random structures.
It will be discussed in this article that averaging over thermal 
fluctuations of heavy atoms provides such a type of disorder (or
quenched) average of physical quantities, like the density 
or the dynamical Green's function of a light atom.
Emphasize is on fermionic system, since the appearence of
Bose-Einstein condensation should not be included in the 
discussion. Moreover, there is a formal mapping of the
heavy fermions to Ising spins (correlated binary alloy).
This provides the opportunity to discuss the statistics
in terms of para-, ferro- and antiferromagnetic Ising states.

The article is organized as follows: In Sect. 2 the model of a
mixture of two types of atoms is discussed for light fermionic
and heavy fermionic or bosonic atoms. The effective distribution
of the heavy atoms is studied in Sect. 3, where it is shown that
for fermionic atoms this distribution is related to a classical
Ising-spin model. The distribution of the Ising spins is investigated
in Sect. 4, using a classical limit (Sect. 4.1) and in a 
strong-coupling expansion (Sect. 4.2). The results are discussed in Sect. 5.

\section{Model}

A mixture of two types of atoms is considered. The atoms are
subject to thermal fluctuations that are treated within a 
grand-canonical ensemble. Moreover, it is assumed that the
atoms are in a magnetic trap such that they are spin polarized.
There is a local repulsive interaction between the two species
but no other interaction of the atoms within the same species. 

$c^\dagger$ ($c$) are creation (annihilation) operators of the
light fermionic atoms, $f^\dagger$ ($f$) are the corresponding
operators of the heavy atoms. The latter can either be fermionic
or bosonic. This gives the formal mapping
\[
^6{\rm Li}\longrightarrow c^\dagger_r, c_r\hskip0.5cm
^{40}{\rm K}\ (^{23}{\rm Na},\ ^{87}{\rm Rb})
\longrightarrow f^\dagger_r, f_r .
\]
\no
The physics of the mixture of atoms is defined by the (asymmetric 
Hubbard) Hamiltonian:
\beq
H= -{\bar t}_c\sum_{\langle r,r'\rangle}c_r^\dagger c_{r'}  
-{\bar t}_f\sum_{\langle r,r'\rangle}f_r^\dagger f_{r'} 
+ \sum_r\Big[-\mu_c c_{r}^\dagger c_{r}-\mu_ff_{r}^\dagger f_{r}
+Uf^\dagger_{r}f_{r}c^\dagger_{r}c_{r}\Big] .
\label{hamilton}
\eeq
An effective interaction among fermionic species is controlled by 
the (repulsive)
Pauli principle. In the case of heavy bosonic atoms the interaction
is neglected. 

The tunneling rate decreases exponentially with the square root of the
mass of the particle. If the $f$ atoms are heavy, the related tunneling rate 
is approximated by ${\bar t}_f\approx 0$. This limiting model is
known as the Falicov-Kimball model 
\cite{falicov69,farkasovsky97,freericks03}.

A grand-canonical ensemble of atoms in an optical lattice is a
situation in which the optical lattice can exchange atoms with a surrounding
atomic cloud. This is given at the temperature $1/\beta$
by the partition function
\[
Z={\rm Tr} e^{-\beta H} .
\]
The real-time Green's function of light atoms describes the
motion of a single light atom, interacting with the atomic mixture.
For the motion from lattice site $r'$ to $r$ during the 
time $t$ it is defined as
\beq
G_c(r,it;r',0)={1\over Z}{\rm Tr}\Big[
e^{-\beta H}c_re^{itH}c^\dagger_{r'}e^{-itH}
\Big].
\label{green2}
\eeq
The density of heavy atoms $n_{f,r}$ and of light atoms 
$n_{c,r}$ can be obtained from the Green's
function in the special case $t=0,r'=r$:
\beq
n_{f,r}={1\over Z}{\rm Tr}\Big[
e^{-\beta H}f_{r}^\dagger f_r\Big],
\hskip0.5cm
n_{c,r}={1\over Z}{\rm Tr}\Big[
e^{-\beta H}c_{r}^\dagger c_r\Big].
\label{density}
\eeq

\section{Ising representation of heavy fermionic atoms}

The quantities of physical interest (density, Green's function etc.) 
are of the form of a trace with respect to $c$ and $f$ atoms:
\[
Tr_{c,f}\left(e^{-\beta H}c_re^{itH}c^\dagger_{r'}e^{-itH}\right).
\]
If ${\bar t}_f=0$ this can also be expressed as
\beq
\sum_{\{ n_r\}}
Tr_{c}\left(e^{-\beta H(\{ n_r\})}
c_re^{itH(\{ n_r\})}c^\dagger_{r'}e^{-itH(\{ n_r\})}\right) ,
\label{sum1}
\eeq
since all operators under the trace are diagonal with respect to the 
number states $|\{n_r\}\rangle$ of the $f$ atoms. Therefore, the Hamiltonian
depends only on the integer numbers $\{ n_r\}$ and the $c$ operators as
\beq
H(\{ n_r\})= -{\bar t}\sum_{\langle r,r'\rangle}c_r^\dagger c_{r'} 
+ \sum_r\left[ n_r(-\mu_f+U c^\dagger_{r}c_{r})
-\mu_c c_{r}^\dagger c_{r}\right]
\equiv-\mu_f\sum_rn_r+H_c(\{ n_r\}) ,
\label{hamilton2}
\eeq
where ${\bar t}_c\equiv{\bar t}$ has been used.
This means that the $f$ density fluctuations $n_r$ have been 
replaced by classical variables:
\[
f_r^\dagger f_r \longrightarrow n_r
\]
which is $n_r=0,1$ in the case of fermions and $n_r=0,1,2,...$
in the case of bosons.
Thus the expression in Eq. (\ref{sum1}) can be written as a sum
over all realizations of $n_r$ as
\[
\sum_{\{ n_r\}}e^{\beta\mu_f\sum_rn_r}
Tr_{c}\left(e^{-\beta H_c(\{ n_r\})}
c_re^{itH_c(\{ n_r\})}c^\dagger_{r'}e^{-itH_c(\{ n_r\})}\right).
\]
$H_c(\{ n_r\})$ is a quadratic form of the $c$ operators. Therefore, it
describes independent spinless fermions which are scattered by heavy
atoms, represented by $n_r$. In particular, the density $n_c$, defined
in Eq. (\ref{density}), reads
\[
n_{c,r}={1\over Z}\sum_{\{ n_r\}}e^{\beta\mu_f\sum_rn_r}
Tr_{c}\left(e^{-\beta H_c(\{ n_r\})}c^\dagger_{r}c_r\right) .
\]
Since $H_c(\{ n_r\})$ is the Hamiltonian of a noninteracting Fermi gas
(it interacts only with the density of heavy atoms),
the trace $Tr_c$ in the partition function can be evaluated as a fermion
determinant \cite{negele88}:
\beq
Z 
=\sum_{\{n_r\}}e^{\beta\mu_f\sum_r n_r}Tr_c
\left(e^{-\beta H_c(\{ n_r\})}\right)
=\sum_{\{n_r\}}e^{\beta\mu_f\sum_r n_r}
{\rm det}[{\bf 1}+e^{-\beta h_c}] ,
\label{partition}
\eeq
where $h_c$ is an $N\times N$ matrix for $N$
lattice sites with matrix elements
\[
(h_c)_{rr'}=-{\hat t}_{rr'}+(Un_r-\mu_c)\delta_{rr'}
\hskip0.5cm
({\hat t}_{rr'}={\bar t}\sum_j\delta_{r',r+e_j}),
\]
where $e_j$ is a lattice unit vector in direction $j$.
A similar calculation gives for the density
\beq
n_{c,r}={1\over Z}\sum_{\{ n_r\}}e^{\beta\mu_f\sum_rn_r}
{\rm det}[{\bf 1}+e^{-\beta h_c}]
[e^{-\beta h_c}({\bf 1}+e^{-\beta h_c})^{-1}]_{rr} .
\label{densitya}
\eeq
In Eq. (\ref{densitya}) there is a non-negative factor
\beq
P(\{ n_r\})={1\over Z}e^{\beta\mu_f\sum_rn_r}
{\rm det}[{\bf 1}+e^{-\beta h_c}]
\label{distr0}
\eeq
which gives $\sum_{\{n_r\}}P(\{ n_r\})=1$ because of Eq. (\ref{partition}).
Thus $P(\{ n_r\})$ is a probability distribution and the expressions
in Eq. (\ref{densitya}) is a quenched average with
respect to this distribution:
\beq
n_{c,r}=\langle
[e^{-\beta h_c}({\bf 1}+e^{-\beta h_c})^{-1}]_{rr}
\rangle_f .
\label{cdensity}
\eeq
The distribution is a realization of correlated disorder and
in the case of fermions it is a correlated binary alloy.
Moreover, the density of heavy atoms 
is given through the distribution $P(\{ n_r\})$ as
\beq
n_f=\langle n_r \rangle_f=\sum_{\{n_r\}}n_rP(\{n_r\}).
\label{fdensity}
\eeq

\section{Approximations for the distribution of heavy fermionic atoms}

Only fermionic mixtures are considered because their
distribution is simpler due to the fact that $n_r$ has only two values. 
The latter implies that $n_r$ can be expressed by Ising spins $S_r=\pm1$ as
\beq
S_r=2n_r-1.
\label{ispin}
\eeq
It is possible to discuss the corresponding distribution in terms of
magnetic Ising states. This will be used in a strong-coupling expansion
of the distribution of the heavy atoms $P(\{S_r\})$ in powers of ${\bar t}/U$.
This is an extension of the strong-coupling approximation for a fermionic
mixture in the symmetric case $\mu_c=\mu_f$ \cite{ates05}.

\subsection{Classical limit: ${\bar t}=0$}

The starting point of the strong-coupling expansion (i.e. the unperturbed
case) is the limit ${\bar t}=0$. This limit is relevant for a mixture in which
both atomic species are so heavy that tunneling can be neglected completely.
The distribution $P(\{n_r\})$ factorizes on the lattice in this case and gives
\beq
P(\{n_r\})\propto \prod_r e^{\beta\mu_f n_r}\left[ 1+e^{\beta(\mu_c-Un_r)}\right] .
\label{distr00}
\eeq
Then the densities of light and heavy atoms are obtained from (\ref{cdensity}) and
(\ref{fdensity}) as
\[
n_c={e^{\beta\mu_c}+e^{\beta(\mu_f+\mu_c-U)}\over
1+e^{\beta\mu_c}+e^{\beta\mu_f}+e^{\beta(\mu_f+\mu_c-U)}
}
\]
\[
n_f={e^{\beta\mu_f}+e^{\beta(\mu_f+\mu_c-U)}\over
1+e^{\beta\mu_c}+e^{\beta\mu_f}+e^{\beta(\mu_f+\mu_c-U)}
} .
\]
The densities of the two species are controlled by their chemical potentials.
At low temperatures (i.e. for $\beta\sim\infty$) they change with $\mu_f$
and $\mu_c$ in a step-like manner (cf. Table 1): The density with the higher
chemical potential is always 1 whereas the density with the lower chemical
potential jumps from 0 to 1 at the value $U$. This means that the repulsive
interaction between the different species prevents the fermionic atoms with the lower
chemical potential to enter the optical lattice unless its chemical potential
is strong enough to overcome the repulsion.  
\begin{table}
\begin{center}
\begin{tabular}{cccc}
 & & $n_c$ & $n_f$ \\
\hline
$\mu_f>\mu_c$: &  &  &  \\
\hline
 & $\mu_c<U$ \vline & 0 & 1 \\
 & $\mu_c>U$ \vline & 1 & 1 \\
\hline
$\mu_f<\mu_c$:  &  &  &  \\
\hline
 & $\mu_f<U$ \vline & 1 & 0 \\
 & $\mu_f>U$ \vline & 1 & 1 \\
\hline
\end{tabular}
\caption[smallcaption]{Mixture of heavy fermionic atoms in the absence of tunneling:
Density of light atoms $n_c$ and of heavy atoms $n_f$ are step-like
functions of the corresponding chemical potentials. The step occurs when the chemical
potential exceeds the repulsive interaction energy $U$.}
\label{table1}
\end{center}
\end{table}

For the symmetric case $\mu_c=\mu_f\equiv\mu$ the densities are equal: $n_c=n_f$.
The behavior is different in comparison with the asymmetric case $\mu_c\ne \mu_f$ 
because of an intermediate
regime for $0<\mu<U$, where both atomic species have to share the optical lattice.
As a consequence, each density then is 1/2:
\beq
n_f=n_c=\cases{
0 & $\mu<0$ \cr
1/2 & $0<\mu<U$ \cr
1 & $\mu>U$ \cr
} .
\label{distr3}
\eeq

\subsection{Strong-coupling (tunneling) expansion: ${\bar t}/U\ll 1$}

The strong-coupling expansion of the distribution function $P(\{S_r\})$
in powers of ${\bar t}/U$ gives an expansion in terms of the Ising
spins \cite{ates05,mix2} (cf. Appendix A):
\[
P(\{ S_r\})\propto \exp\left(
E_1\sum_rS_r+{E_2\over2}\sum_{r,r'}S_rS_{r'}
\right)\prod_r\left[
1+e^{\beta(\mu_c-U(1+S_r)/2)}
\right]
\]
with coefficients
\[
E_1=
{\beta^2{\bar t}^2\over 8}\left[
{e^{\beta\nu-\beta g}\over(1+e^{\beta\nu-\beta g})^2}
-{e^{\beta\nu+\beta g}\over(1+e^{\beta\nu+\beta g})^2}
\right]
\]
and
\[
E_2=
\left\{
{\beta^2{\bar t}^2\over 8}\left[
{e^{\beta\nu-\beta g}\over(1+e^{\beta\nu-\beta g})^2}
+{e^{\beta\nu+\beta g}\over(1+e^{\beta\nu+\beta g})^2}
\right]
-{\beta\over 4g}{e^{\beta\nu}\sinh(\beta g)\over
1+e^{2\beta\nu}+2e^{\beta\nu}\cosh(\beta g)}
\right\}.
\]
These coefficients simplify substantially in the low-temperature regime 
($\beta\sim\infty$):
\[
E_1\sim0,\ \ \ E_{2}\sim \cases{
-\beta{\bar t}^2/4U
& for $0<\mu_c<U$ \cr
0 & otherwise \cr
}
\]
which gives three different regimes for the distribution:
\beq
P(\{ S_r\})\propto\cases{
\exp\left[\beta{\mu_f\over 2}\sum_r S_r
\right] & $\mu_c<0$ \cr
\exp\left[
\beta\left(
{\mu_f-\mu_c\over2}\sum_rS_r-{{\bar t}^2\over4U}
\sum_{<r,r'>}S_rS_{r'}\right)\right] & $0<\mu_c<U$ \cr
\exp\left[\beta{(\mu_f-U)\over 2}\sum_r S_r\right]
& $U<\mu_c$ \cr
} .
\label{scdistr}
\eeq
This distribution is a classical Ising model with a magnetic field term
$h\sum_rS_r$ and an antiferromagnetic spin-spin interaction
$-J\sum_{<r,r'>}S_rS_{r'}$ with $J>0$. The latter is absent in the regimes
with $\mu_c<0$ and $\mu_c>U$. These regimes are only controlled by an
effective magnetic field $\mu_f/2$ and $(\mu_f-U)/2$, respectively: A
positive field yields a positive Ising spin, implying that the optical lattice
is fully occupied by heavy atoms. A negative field yields a negative spin,
implying the absence of heavy atoms in the optical lattice. 
However, there is no real competition between the two atomic species. 
In the intermediate regime $0<\mu_c<U$, on the other hand, there is competition
due to the appearence of the linear and the bilinear spin terms:
\beq
H_I={\mu_f-\mu_c\over2}\sum_rS_r-{{\bar t}^2\over4U}\sum_{<r,r'>}S_rS_{r'} .
\label{iham}
\eeq
The linear term favors a homogeneous distribution of heavy atoms whereas the
antiferromagnetic bilinear term favors a staggered distribution of heavy atoms
(i.e. a site that is occupied by a heavy atom has nearest neighbor sites without 
heavy atoms).
The latter is caused by the tunneling of the light atoms, since it is proportional
to the square of tunneling rate of the light atoms ${\bar t}$. A simple mean-field
calculation reveals that there is a first-order phase transition between the
homogeneous phase and the staggered phase. Thermal fluctuations lead to the 
competition of homogeneous and staggered clusters \cite{ates05,mix2}. A typical
realization of the distribution of heavy atoms, created by a Monte-Carlo simulation,
is shown in Fig. 1. 

At $\mu_f=\mu_c=U/2$ the distribution in Eq. (\ref{distr0}) has a 
global Ising-spin flip symmetry
and is related to a half-filled system. This means that half of the
sites of the optical lattice are occupied by heavy atoms. In this case there is
a continuous phase transition from antiferro- to paramagnetic order
if the temperature is increased.

\section{Discussion}

The complex interplay between light and heavy fermionic atoms in mixtures results
in a correlated distribution of the heavy atoms, given by the distribution density
of Eq. (\ref{distr0}). The correlations are a consequence of the tunneling processes
of the light atoms, i.e. they are due to quantum effects. If quantum tunneling is
suppressed (e.g., by high barriers in the optical lattice and/or by using heavy atoms),
the distribution of the heavy atoms looses the correlations, as discussed in Sect. 
4.1. This is a consequence of our simple model where only local interatomic 
interactions are considered. In the presence of a nonlocal interaction between 
the atoms correlations
would also exist in the absence of quantum tunneling. However, it is unlikely that
nonlocal interactions play a crucial role for neutral atomic mixtures in an optical
lattice because it is dominated by $s$-wave scattering.

The correlations of the local density fluctuations of the heavy atoms lead to
different phases and phase transitions. A perturbation expansion for weak
tunneling (i.e. tunneling ${\bar t}$ is weak in comparison with the interatomic 
interaction $U$), presented in Sect. 4.2, has revealed that the heavy atoms are
homogeneously distributed with one atom per optical lattice site 
if their chemical potentials $\mu_c$ and $\mu_f$ are larger than the interatomic interaction
$U$. According to the Ising Hamiltonian $H_I$ of Eq. (\ref{iham}), which describes
the distribution for $0<\mu_c<U$, a homogeneous distribution of heavy atoms 
also exists if $\mu_f-\mu_c$ is large in comparison with
${\bar t}^2/4U$. On the other hand, for $|\mu_f-\mu_c|$ small in comparison with
${\bar t}^2/4U$ the heavy atoms are arranged in a staggered order, similar to a
charge-density wave in solid-state physics \cite{kittel87} with $n_f\approx 0.5$.
For $\mu_f-\mu_c\ll -{\bar t}^2/4U$ heavy atoms are pushed out of the optical lattice.
These results indicate
that quantum tunneling causes an effective repulsive nonlocal interaction between
the heavy atoms. The Ising Hamiltonian $H_I$ yields a first order transition
from the homogeneous to the staggered distribution for a decreasing $\mu_f-\mu_c>0$.
The associated coexistence of clusters with both types of order (see also Fig. 1)
is similar to phase separation, discussed in the solid-state literature
\cite{littlewood,dagotto}. 
Furthermore, there is a second order transition from the staggered distribution at
low temperatures to a disordered distribution at higher temperatures.
In experiments the phase transitions are easily accessible by a
variation of the tunneling rate ${\bar t}$ in the optical lattice.

The symmetric situation $\mu_f=\mu_c$, where the optical lattice does not
distinguish between the heavy and the light atoms energetically, there is an
additional type of competition between heavy and light atoms even in the
absence of quantum tunneling: although there is no order in the distribution
of heavy atoms, the sites of the optical lattice can be randomly occupied
either by light or by heavy atoms, provided the repulsion is strong enough
(i.e. for $0<\mu_c=\mu_f<U$). Consequently, the average density is $n_c=n_f=0.5$
(cf. Eq. (\ref{distr3})). The total symmetric case $\mu_f=\mu_c$ and 
${\bar t}_f= {\bar t}_c$ is the Hubbard model. At half-filling the strong-coupling
expansion leads to an antiferromagnetic spin-1/2 Heisenberg model \cite{fulde93}.
The ground state of this model is a staggered (Ne\'el) state.
If the spin is associated with two atomic species, the ground state of the
atomic system is an alternating arrangement of the two type of atoms in the 
optical lattice.

Light atoms are also affected by the order of the heavy atoms 
in the mixture, since they are scattered by the ensemble of heavy atoms. 
As a result, they can propagate (if heavy atoms are ordered, e.g., at
low temperatures),
they can diffuse (if heavy atoms are weakly disordered),
or they are localized (if heavy atoms are strongly disordered).
Due to correlations of the distribution of heavy atoms,
a gap in the density of states of the light atoms can
be opened \cite{ates05,mix2}. This leads to an incompressible state
of light atoms, where diffusion is absent. For disorder of
heavy atoms (i.e. in the regime of competing clusters or at 
higher temperatures) localization of light atoms can take place.

\section{Conclusions}

There is a complex interplay between light and heavy
fermionic atoms, where the latter can be described by classical 
Ising spin. The nonlocal interaction between the heavy atoms is caused
by quantum tunneling of the light atoms.
This is given by an effective Ising model with a symmetry-breaking
magnetic field and an antiferromagnetic spin-spin coupling,
at least in the regime of strong coupling between the two
atomic species. It implies a competition between 
homogeneous and staggered order in the distribution of heavy atoms.
There is a first order transition from homogeneous to staggered order
and a second order transition from staggered order to a disordered
distribution of heavy atoms.
\vskip0.5cm
\no
Acknowledgement: I am grateful to S. Fialko for providing Fig. 1.

\section*{Appendix A}

Using $\nu=\mu_c-U/2$ and $g=U/2$ we can write
the determinant in equation (\ref{distr0}) as a power series of
${\bar t}$ (i.e. in the strong-coupling regime):
\[
\exp\left\{\Tr\left[
\ln\left({\bf 1}+e^{\beta(\nu+{\hat t}-gS)}\right)
\right]\right\}
\approx
\det \left(G_1^{-1}\right)\exp\left[
\Tr(G_1 D)-{1\over2}\Tr(G_1 D G_1 D)
\right]
\]
\[
{\rm with} \quad G_1^{-1}={\bf 1}+e^{\beta(\nu-gS)} \ \ \ {\rm and}\ \
D=e^{\beta(\nu+{\hat t}-gS)} -e^{\beta(\nu-gS)}.
\]
A lenghty but straightforward calculation gives with
$A_r=\nu-gS_r$ the relation
\[
\Tr(G_1 D)-{1\over2}\Tr(G_1 D G_1 D)\approx{\beta\over 2}
\sum_{r,r'}
{e^{\beta A_r}-e^{\beta A_{r'}}\over A_r-A_{r'}}
{{\bar t}^2\over(1+e^{\beta A_r})
(1+e^{\beta A_{r'}})} .
\]
Since there are only values $S_r=-1,1$, the term
\[
E(S_r,S_{r'})={e^{\beta A_r}-e^{\beta A_{r'}}\over A_r-A_{r'}}
{{\bar t}^2\over(1+e^{\beta A_r})(1+e^{\beta A_{r'}})}
\]
can also be expressed as a quadratic form with respect to the
Ising spins:
\[
E(S_r,S_{r'})=E_0+E_1(S_r+S_{r'})+E_2S_rS_{r'}
\]
with
\begin{eqnarray}
E_0 & = & {1\over4}\left[E(1,1)+E(-1,-1)+2E(1,-1)\right]\nonumber\\
& = & 
{\bar t}^2\left\{
{\beta^2\over 8}\left[
{e^{\beta\nu-\beta g}\over(1+e^{\beta\nu-\beta g})^2}
+{e^{\beta\nu+\beta g}\over(1+e^{\beta\nu+\beta g})^2}
\right]
+{\beta\over 4g}{e^{\beta\nu}\sinh(\beta g)\over
1+e^{2\beta\nu}+2e^{\beta\nu}\cosh(\beta g)}
\right\},\nonumber
\end{eqnarray}
\begin{eqnarray}
E_2 & = & {1\over4}\left[E(1,1)+E(-1,-1)-2E(1,-1)\right]\nonumber\\
& = & 
{\bar t}^2\left\{
{\beta^2\over 8}\left[
{e^{\beta\nu-\beta g}\over(1+e^{\beta\nu-\beta g})^2}
+{e^{\beta\nu+\beta g}\over(1+e^{\beta\nu+\beta g})^2}
\right]
-{\beta\over 4g}{e^{\beta\nu}\sinh(\beta g)\over
1+e^{2\beta\nu}+2e^{\beta\nu}\cosh(\beta g)}
\right\},\nonumber
\end{eqnarray}
and
\[
E_1={1\over4}[E(1,1)-E(-1,-1)]
={\beta^2{\bar t}^2\over 8}\left[
{e^{\beta\nu-\beta g}\over(1+e^{\beta\nu-\beta g})^2}
-{e^{\beta\nu+\beta g}\over(1+e^{\beta\nu+\beta g})^2}
\right].
\]
A further contribution comes from 
\[
\det \left(G_1^{-1}\right)=\prod_r\left[1+e^{\beta(\nu-gS_r)}\right] .
\]

\begin{figure}
\begin{center}
\includegraphics[width=0.3\textwidth]{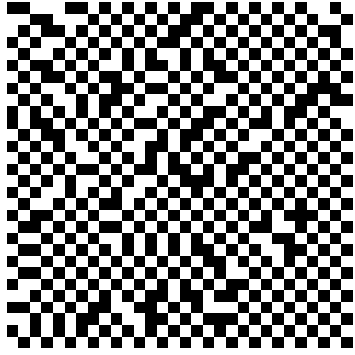}
\end{center}
\caption{Competition of homogeneous and staggered clusters of heavy
fermionic atoms (indicated as dark squares): a typical realization
from the distribution in Eq. (\ref{distr0}) at $\mu_f=\mu_c=U/2$.}
\end{figure}

\end{document}